\documentclass[9pt]{article}
\usepackage[left=1.5cm, right=1.5cm, bottom=1.5cm, top=1.5cm]{geometry}
\usepackage{amsmath}
\usepackage{amssymb}
\usepackage{graphicx}
\usepackage{multicol}
\usepackage{float}
\usepackage{verbatim}
\usepackage{blindtext}
\usepackage{tcolorbox}
\usepackage{pgfplots}
\usepackage{epsfig}
\usepackage{epstopdf}

\title{Anomalous cumulative inertia in human behaviour}
\author{Stage, H. B.$^*$ and Fedotov, S. P.}
\date{}

\begin{document}
\maketitle

\begin{multicols}{2}

\textbf{Human behaviour is dictated by past experiences via cumulative inertia (CI): the longer a certain behaviour has been going on, the less likely changes becomes. This is a well-known sociological phenomenon observed in employment, residence, addiction, criminal activity, wars, etc. Fundamentally, these all exhibit a growing resistance to change over time.
However, quantifying the strength of this inertia is an ongoing challenge. Here we uncover anomalous cumulative inertia (ACI), ubiquitous across human behavioural patterns, with a much stronger memory dependence than previously anticipated. The behaviours undergo substantially stronger inertia, invalidating classical predictions for recovery, reconciliation, or rehabilitation times. We propose alternative models for predictions of continued anomalous behaviour, and provide means of identifying whether such behaviour is present. The result is a paradigm shift in our understanding of human activity from burstiness to inertia.
Our results demonstrate how non-equilibrium models using fractional calculus aptly describe resistance to behavioural change, and produce novel predictions for e.g.\ rehabilitation of convicted individuals. The presence of anomalous cumulative inertia qualitatively affects the predictions which can be made for behavioural change, and thus forecasts which of these are more or less changeable. These findings are critically important for e.g.\ recidivism studies and public policy making, by determining more successful intervention strategies and populations where these interventions are most likely to succeed.}

\textit{Cumulative inertia} (CI) is a fundamental property of human behaviour, whether we are conscious of it or not \cite{first-inertia}. It is the statement \textit{that the longer we retain a job, own a home, or belong to a certain political group, the less likely we are to change those circumstances}. 
Consider for example tenured academics: with time, the individuals increase in seniority and salary rank, develops a better working relationship with colleagues, etc. It so becomes increasingly unlikely that they will change institution. Once at full professorial level academics are essentially \textit{trapped} in their current position.
This has been observed across a multitude of human employment contexts \cite{employmentdata, jobs, hetero}. CI is also historically well-documented in human conflicts. A war or strike which has gone on for 3 months is far more likely to continue for another three months than a confrontation which only started 2 weeks ago (both parties have had greater opportunity to become entrenched in their beliefs, reducing chances of a compromise) \cite{jobs, war-data}.
CI is prevalent in online streaming; the longer our streaming session, the more likely we are to continue \cite{netflix}; an effect also seen in television consumption \cite{tv}. The same is true for correspondence where the likelihood of responding decreases with the time since receiving the message \cite{letters, emails}, and in political activity where inveterate members are less likely to leave than new members \cite{pewresearch}.

CI was popularised during studies of human movement and migration \cite{Myers67, Gordon95, prospect, humanmove}, where local attachments were observed to lead to reluctant movers. It has recently experienced a resurgence with new empirical support for postulated psychological interpretations of the phenomenon \cite{prospect, collation, prl-1}. In the seminal work of Kahnemann and Tversky, the \textit{endowment effect} was put forth as an explanation of this aversion to change. As we experience our home and neighbourhood, a set of positives and negatives is acquired (amenities, commute, neighbours, etc.) which forms an endowment. Changing means the loss of this endowment and the risk that the gains of changing will not outweigh this. The result is an aversion to (risk of) loss which biases the decision \cite{kahn-book}.

\begin{figure}[H]
\centering
\includegraphics[scale=1]{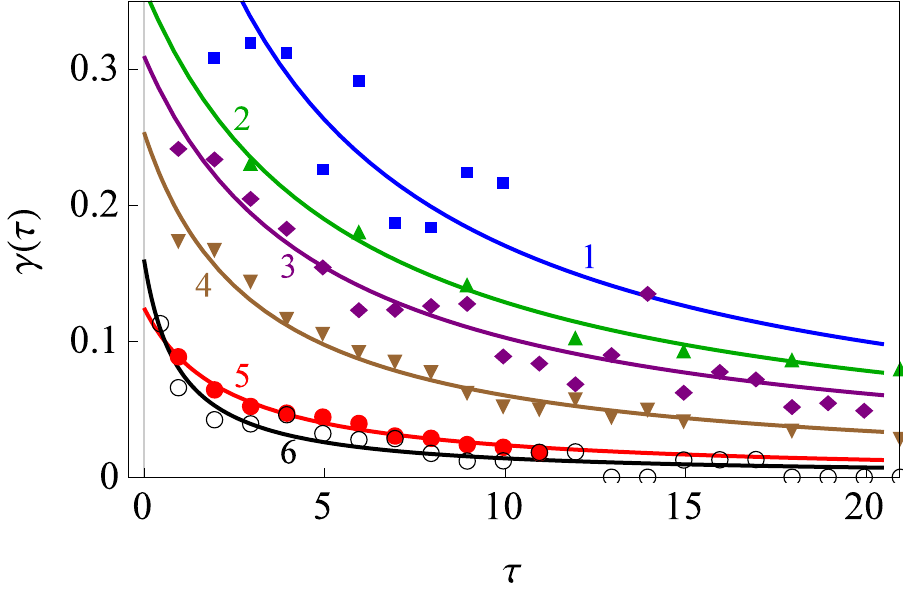}
\caption{\label{fig: emp-rates} Decreasing empirical rates of change $\gamma$ as a function of time $\tau$, fitted to the curve $\mu/(\tau+\tau_0)$. Changes in these human behaviours (over days D, months M, or years Y) depend on rate parameters $\mu$ and time scales $\tau_0$. Fitting details in SI, part 2. In descending order, 1-- war durations (Y, $\mu=2.42$) \cite{jobs}, 2-- factory worker employment (M, $\mu=2.01$) \cite{jobs}, 3-- strike durations (D, $\mu=1.54$) \cite{jobs}, 4-- steel workers employment (M, $\mu=0.79$) \cite{jobs}, 5-- Finnish movement post-graduation (Y, $\mu=0.29$) \cite{fin-grads}, and 6-- recidivism of parolees (M, $\mu=0.15$) \cite{crime2}.}
\end{figure}

CI can be expressed in terms of a rate of change $\gamma(\tau)$ which decreases with time $\tau$ \cite{jobs, hetero}. The expression $\gamma(\tau)\Delta\tau$ is the conditional probability of e.g.\ being employed for a duration in the interval $[\tau,\tau+\Delta\tau]$, given the employment has already lasted a duration $\tau$. On dimensional grounds, let the rate of change take the form $\gamma(\tau)= \mu/(\tau+\tau_0)$, such that with increasing time $\tau$ spent, the smaller the rate of change (rate parameter and time scale $\mu,\tau_0>0$). Empirical evidence for this dependence is illustrated in Figure \ref{fig: emp-rates}. If a time $t$ has already been spent in a job ($\tau\geq t$), the \textit{mean remaining time} $\left<T|_{\tau\geq t}\right> = (\tau_0+t)/(\mu-1)$ exhibits inertia because it increases with $t$ (see \cite{jobs} and SI, part 1) This expression holds for $\mu>1$, a key property of CI.

\begin{figure}[H]
\centering
\includegraphics[scale=1]{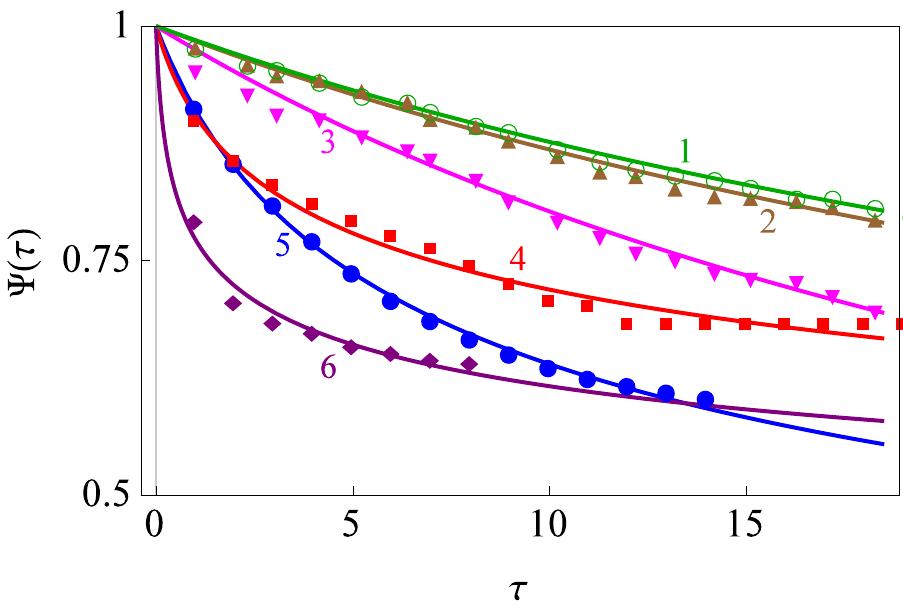}
\caption{\label{fig: emp-survival} The curves display the survival probabilities $\Psi(\tau)=\tau_0^\mu/(\tau+\tau_0)^\mu$ of individuals remaining in their current state (e.g.\ home or behavioural pattern) for a certain duration $\tau$ (in months M, or years Y). Fitting details in SI, part 2. In descending order, 1-- recidivism of non-drug parolees (M, $\mu=0.44$), 2-- recidivism of drug parolees (M, $\mu=0.48$), 3-- recidivism non-drug ex-prisoners (M, $\mu=0.71$) \cite{crime3}, 4-- post-graduation moves in Germany (Y, $\mu=0.13$) \cite{ger-grads}, 5-- post-graduation moves in Finland (Y, $\mu=0.28$) \cite{fin-grads}, and 6-- net recidivism of Spanish convicts (Y, $\mu=0.10$) \cite{crime1}. The population changes can be obtained by multiplying the initial population by $\Psi(\tau)$.}
\end{figure}

\begin{figure}[H]
\centering
\includegraphics[scale=1]{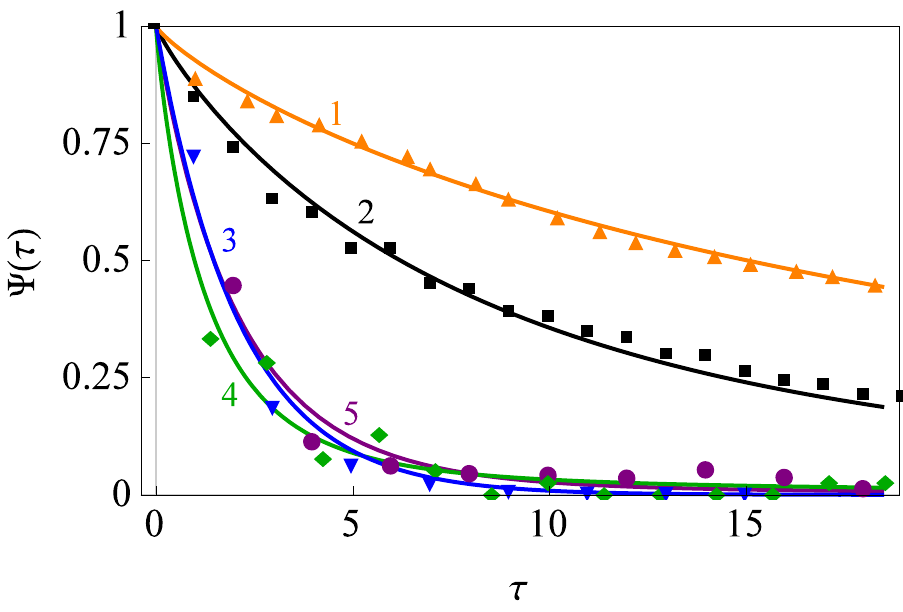}
\caption{\label{fig: emp-ml-survival} The curves display the survival probabilities $\Psi(\tau)=E_\mu\left(-(\tau/\tau_0)^\mu\right)$ of individuals remaining in their current state (e.g.\ home or behavioural pattern) for a certain duration $\tau$ (in fortnights F, months M, half-years H, or years Y), where $E_\mu$ is the Mittag-Leffler function. Fitting details in SI, part 2. In descending order, 1-- recidivism of drug ex-prisoners (M, $\mu=0.84$) \cite{crime3}, 2-- human residence (Y, $\mu=0.89$) \cite{Anily99}, 3-- Syrian refugees (H, $\mu=1.00$) \cite{syria}, and 4-- relapse durations of alcoholics (F, $\mu=0.89$) \cite{alcoholics}. 5-- duration likelihoods of inter-state conflicts (F, $\mu=0.96$) \cite{war-data}.}
\end{figure}

In this work we uncover a hitherto untreated trend in human behaviour which we call \textit{anomalous} cumulative inertia (ACI), supported by empirical evidence. Crucially, the predictions from standard cumulative inertia regarding the mean remaining time break down due to a smaller rate parameter $\mu<1$. Instead, anomalous systems possess a much stronger inertia than standard CI, to the extent that trapping (or no change) is a highly likely scenario. Surprisingly, it is also rather common. ACI concerns cases in which additional effort is required in order to bring about change. Examples include living in an affluent/pleasant area and moving \cite{milwaukeedata}, leaving a stable well-paying job, or our activities on the internet \cite{internet, internet1, Liu10}, including `clickbait' websites which do their utmost to keep a user on the site for as long as possible \cite{clickbait}. 
A large body of literature examines bursty activity followed by long periods of inactivity for related examples, but do not address anomalous effects \cite{bar1, bar2, bar3}.\newline
A hallmark of anomalous cumulative inertia is that classical treatments of cumulative inertia break down due to the long-term memory effects in place. Consequently, predictions pertaining to the mean remaining time $\left<T|_{\tau\geq t}\right>$ no longer apply; the mean has ceased to exist. The consequences of ACI are pervasive beyond the breakdown of simple quantities like the mean remaining time. They demand a new family of mathematical treatments: fractional calculus valid when the rate parameter $\mu<1$ as tested by empirical data in Figures \ref{fig: emp-survival} and \ref{fig: emp-ml-survival}. Fractional calculus was popularised in physics as a means of describing anomalous transport \cite{guide, mee-book, kla-book, sok-book}.

Before addressing this point in greater detail, we first motivate the rate of change $\gamma(\tau)=\mu/(\tau+\tau_0)$, which describes both CI and ACI depending on the value of rate parameter $\mu$. Figure \ref{fig: emp-rates} collates data for rates of change observed in a variety of human behaviours. These include the rates at which wars (1820-1949) or British strikes (in 1965) end, depending on their duration thus far. We also analyse the rates at which salaried individuals change jobs (1944-1959), and the rate at which a convicted individual in Illinois, U.S.A.\ granted parole will return to crime. Finally, we consider the rate at which Finnish individuals leave their university city after graduating.
While the curves look deceptively similar with only small variations in the $\mu-$values, Figure \ref{fig: emp-rates} contains two different classes of behaviour. We observe both CI with $\mu>1$ and ACI with $\mu<1$. We shall demonstrate how this innocuous change in $\mu$ leads to significantly altered predictions and expectations of past and future behaviour. Physically, this can be viewed as a non-equilibrium phase transition at $\mu=1$.

In Figure \ref{fig: emp-survival} we provide further empirical support for the presence of ACI. We consider the survival probabilities $\Psi(\tau)$ of convicted individuals in the U.S.A. and Spain not committing another crime for a given duration. The American data distinguishes between individuals sent to prison or given parole, as well as if the crime is drug-related or not. We also consider the probabilities of Finnish and German individuals remaining in their university city after graduation for some time.
The plotted functional form of $\Psi(\tau)$ follows directly from the rate $\gamma(\tau)$ of Figure \ref{fig: emp-rates} via $\Psi(\tau)=\exp\left(-\int_0^\tau\gamma(u)du\right)$, containing heavy tails of the survival probability with $\mu<1$. It is these heavy tails which lead to the necessity of fractional calculus \cite{sok-book}. 

The tails of $\Psi(\tau)$ plotted in Figure \ref{fig: emp-survival} are equivalent to those arising from the solutions to fractional equations in \eqref{eq: rl-der} and \eqref{eq: cap-der}. The resulting Mittag-Leffler function $\Psi(\tau)=E_\mu\left(-(t/\tau_0)^\mu\right)$ is an ideal fitting candidates for empirical cases when the tails dominate the dynamics, capable of describing systems with strong long-time effects.
This is evidenced in Figure \ref{fig: emp-ml-survival}, where we consider the probability of an American individual imprisoned for a drug-related crime to recommit another offence, given a certain duration passed without convictions. We also consider the probabilities for different durations individuals will remain in their current residence, or in the EU if seeking asylum. Furthermore, we consider the times passed until alcoholic individuals will start drinking again, and the durations of militarised inter-state conflicts (1816-2010).

\begin{tcolorbox}[width=.45\textwidth,colback={blue!10!white},title={Fractional Calculus and ACI},colbacktitle=blue!20!white,coltitle=black]
Let us consider the survival probability of employees $\Psi(t)$ in a company which exhibits ACI. The associated strong inertia leads to the fractional equation for $\Psi(t)$
\begin{equation}
\frac{d\Psi}{dt}=-\frac{1}{\tau_0^\mu}\ _0\mathcal{D}_t^{1-\mu}\Psi(t),
\label{eq: rl-der}
\end{equation}
with the Riemann-Liouville fractional derivative defined as $_0\mathcal{D}_t^{1-\mu}\Psi(t)=\frac{1}{\Gamma(\mu)}\frac{d}{dt}\int_0^t\Psi(t-u)/u^{1-\mu}du$ \cite{rl-operator}. This clearly has a strong memory dependence. An other formulation of the form
\begin{equation}
\tau_0^\mu\frac{d^\mu\Psi}{dt^\mu}=-\Psi(t),
\label{eq: cap-der}
\end{equation}
using the fractional Caputo derivative \cite{caputo} also exists. 
The solution to \eqref{eq: rl-der} and \eqref{eq: cap-der} is given by the  Mittag-Leffler function $\Psi(t)= E_\mu\left(-(t/\tau_0)^\mu\right)$, which is illustrated in Figure \ref{fig: emp-ml-survival}.
\end{tcolorbox}

\begin{figure}[H]
\begin{tikzpicture}
\node (img1) {\includegraphics[width=0.49\textwidth]{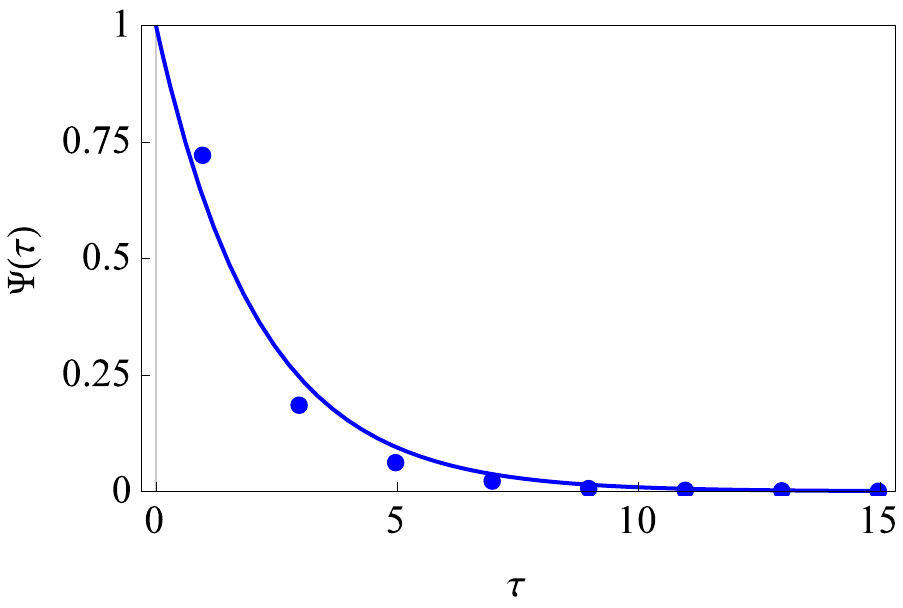}};
  \node (maps) at (img1) [yshift=.8cm,xshift=1.45cm] {\includegraphics[width=5.5cm]{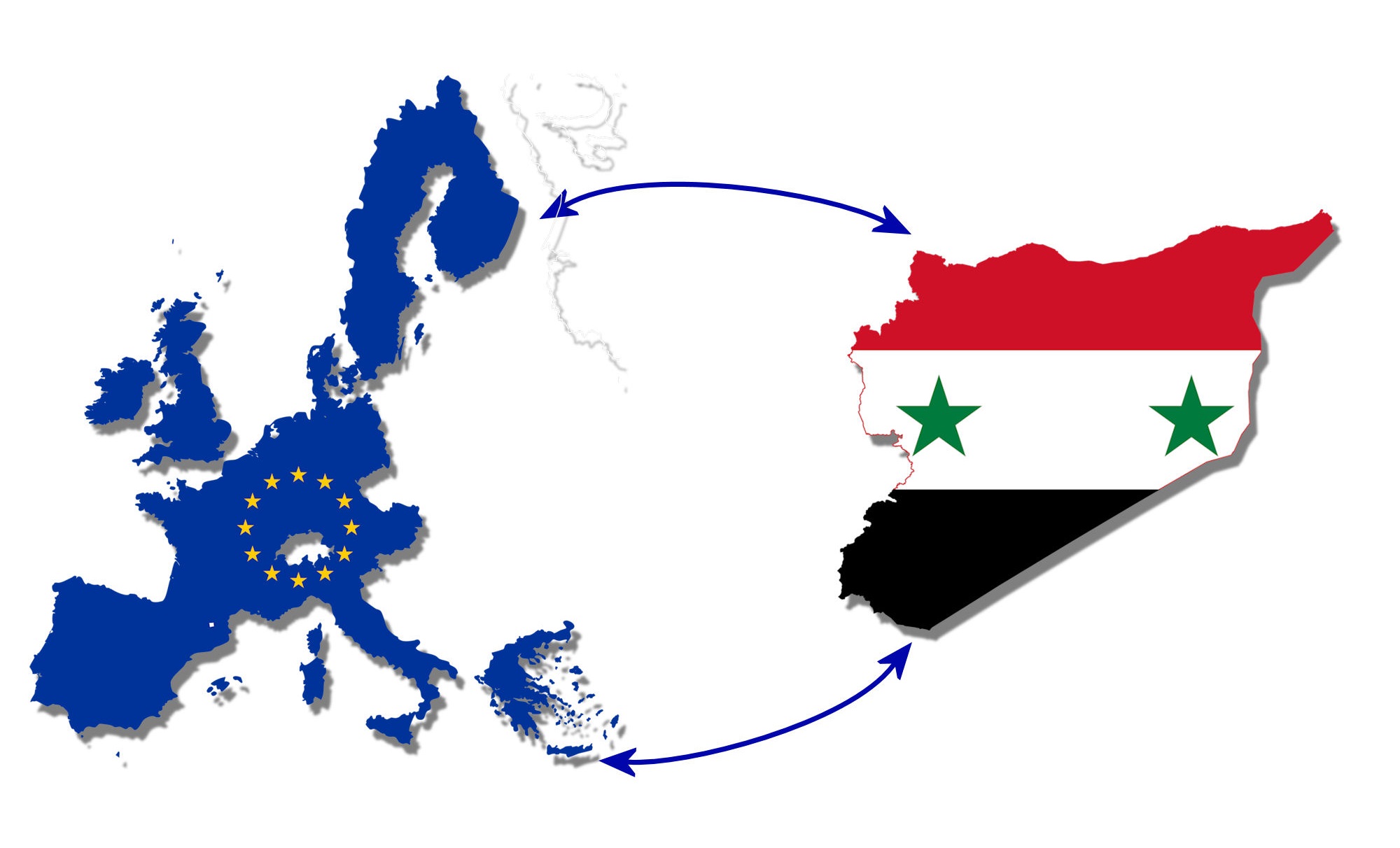}};
\end{tikzpicture}
\caption{\label{fig: diagram} Survival probabilities $\Psi(\tau)$ of Syrian refugees (2008-present) remaining in the EU, estimated via asylum applications and deportations. The residence durations $\tau$ are measured in 6-month intervals. During times of unrest, the rate of leaving Syria is approximately constant, whereas the time spent in the EU makes the likelihood of re-settling there higher. With a value of $\mu\approx1$, the long-term living situation of refugees is precarious; small changes could drastically affect whether individuals remain for longer times or if they are deported. This is a critical example of a two-state system wherein accounting for the fractional dynamics of the individuals is essential. We find that the growth of refugee populations (assuming no change in influx) is governed by heavy-tailed dynamics.}
\end{figure}

The empirical support for ACI and fractional calculus is provided in Figure \ref{fig: emp-survival} and \ref{fig: emp-ml-survival}. The importance of fractional calculus becomes clear when one considers the competition across a network of companies or universities \cite{prl-1}.
Highly reputable or prestigious companies can reasonably be modelled by ACI, and out-compete locations with classical CI. Worker allocation will then be dominated by companies with ACI. A heuristic interpretation is the well-established notion of `brain-drain' between different countries, or international migration between regions of significantly different living standards.
Naturally the rates of leaving are representative of the different locations, but the question is how this affects the overall distribution of individuals. 
If one considers two competing companies, one of which exhibits ACI, an equilibrium distribution of the workers will never be reached \cite{prl-1}. Instead, most workers will be found in the state with ACI, which we emphasize is different from an absorbing state. Individuals may still change, but overall are trapped in the anomalous state.
In Figure \ref{fig: diagram} we illustrate this by the trapping of arriving Syrian refugees in the EU.

We now demonstrate the breakdown of standard techniques for modelling and data analysis concerning behaviours in systems subject to ACI. In what follows we illustrate this with the example of employees in a company founded at time $t=0$. If each employee leaving their job is replaced immediately, every job generates a renewal process \cite{feller}. Such a renewal process involving \eqref{eq: cap-der}, is known as a fractional Poisson process \cite{frac-pois}.
This can be gauged via the structural population density $n(t,\tau)$: we consider not only the number of individuals in the company at time $t$, but also the time $\tau$ they have held their current job. A similar analysis can be done for residence durations. Classical techniques, based on equilibrium distributions, assume that the distribution of durations is independent of time $t$ (see \eqref{eq: eq-dist}) \cite{Anily99}. However, the strong memory of ACI invalidates the assumptions of these techniques, resulting in fundamentally altered, time-dependent employment or residence time distributions (see \eqref{eq: struct-anom}).
For CI, the equilibrium density $n_{eq}(\tau)$ is a decaying function of duration time $\tau$. However, a hallmark of ACI is the presence of long memory effects, leading to a drastically different U-shaped form of $n(t,\tau)$ instead.
This is a clear example of the effects of anomalous behaviour: the inertia is much stronger, allowing for a large probability of individuals who do not change. 

\begin{tcolorbox}[width=.45\textwidth,colback={blue!10!white},title={Properties and Predictions of ACI},colbacktitle=blue!20!white,coltitle=black]
If employment at a company follows CI, then the distribution of employment durations is a decreasing function of duration $\tau$:
\begin{equation}
n_{eq}(\tau)=\frac{\Psi(\tau)}{\left<T\right>},
\label{eq: eq-dist}
\end{equation}
\vspace*{2.1cm}
\begin{center}
\smash{\raisebox{0pt}{\includegraphics[width=6.5cm]{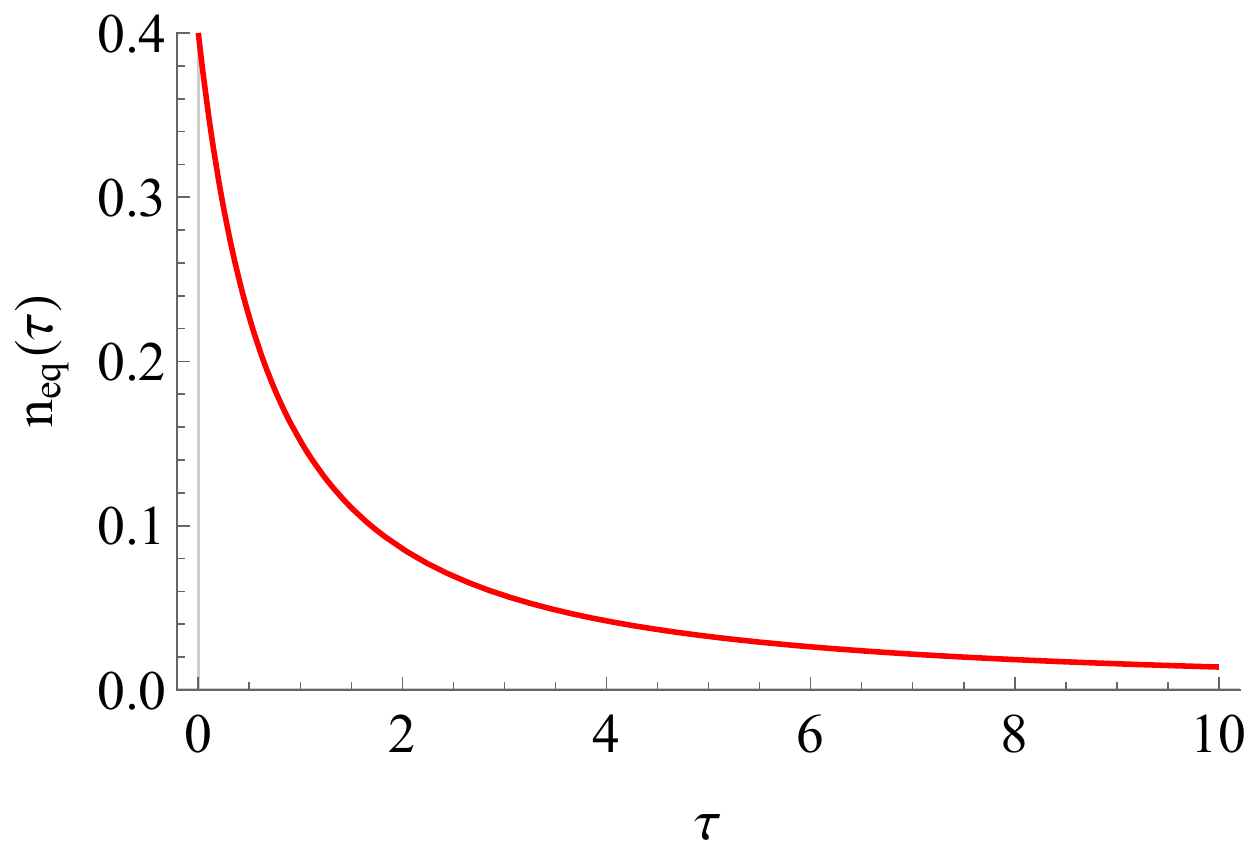}}}
\vspace*{-0.7cm}
\end{center}
plotted here for $\mu=1.4,\ \tau_0=1$.
For ACI, the distribution is radically changed to a U-shape:
\begin{equation}
n(t,\tau)=\frac{A}{\tau^\mu(t-\tau)^{1-\mu}},
\label{eq: struct-anom}
\end{equation}
\vspace*{2.2cm}
\begin{center}
\smash{\raisebox{0pt}{\includegraphics[width=6.5cm]{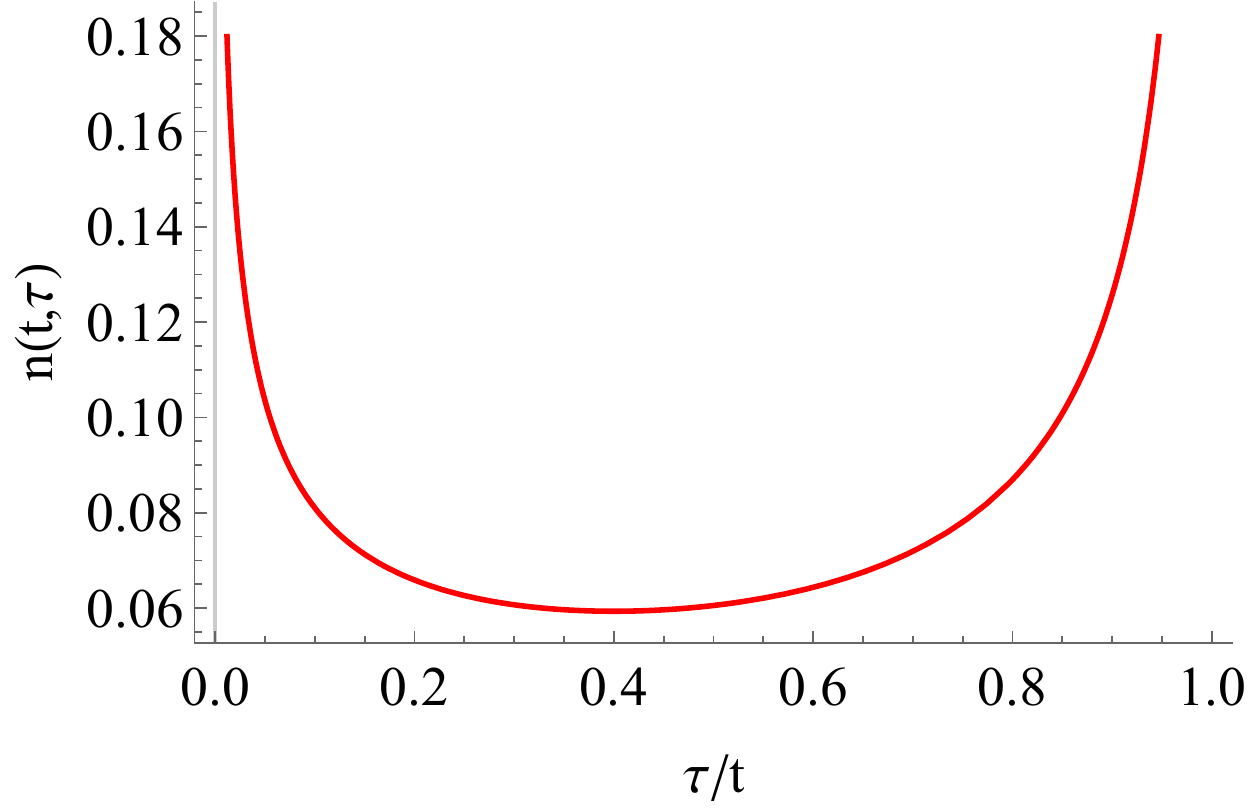}}}
\vspace*{-0.7cm}
\end{center}
with high probabilities of short $\tau\ll t$ and very long $\tau\sim t$ durations (see SI, part 3 and \cite{feller}). The constant $A=[\Gamma(\mu)\Gamma(1-\mu)]^{-1}$, as is plotted for $\mu=0.4,\ t=10$. Crucially, the presence of $t$ in \eqref{eq: struct-anom} implies a strong memory dependence.
The population density $\phi(t,x)$ of workers remaining for an additional duration $x$ is
\begin{equation}
\phi(t,x)=\frac{A t^\mu}{(t+x)x^\mu},
\label{eq: aci-pred}
\end{equation}
which is also time-dependent (see \cite{feller,sok-book}). The constant $A$ takes the same value as above. Conversely, for CI $\phi(t,x)$ takes the form of \eqref{eq: eq-dist}.
\end{tcolorbox}

The high probability of durations $\tau\sim t$ in \eqref{eq: struct-anom} are indicative of the long-time persistence of systems with ACI.
The presence of this U-shape is thus indicative of the high likelihood of continuation of the current persistent behaviour.
Consequently, ACI has profound implications for the long-term expectations of human behaviour. This can be seen in the time-dependent heavy tails of \eqref{eq: aci-pred} which give high probabilities for the continuation of the anomalous behaviour.

If \textit{recidivism}, the likelihood of a criminal to reoffend, is subject to inertia \cite{crime1, crime2, crime3}, we predict qualitatively different outcomes if the inertia is anomalous. If the inertia is `classical' with $\mu>1$, the average individual may attempt behavioural change for some time, but will almost inevitably return to crime \cite{crime1,crime2,crime3}. This can be seen from the finite mean remaining time \cite{jobs}.
However, if the recidivism inertia is anomalous with $\mu<1$, the average individual is unlikely to succeed in changing behaviour.
In other words, the expected recidivism time exceeds the human life span, though the median time may still be evaluated (see SI, part 1).
Any criminal activity can be gauged via the conditional survival probability $\Psi(\tau,x):$ given no criminal offence has taken place for a duration $\tau$, what is the probability of this pattern continuing for an additional duration $x$? For both CI and ACI this is given by $\Psi(\tau,x)=(1+x/\tau)^{-\mu}$, as derived in the SI, part 1 \cite{jobs}.
A successful strategy for criminal intervention would thus identify populations with $\mu$-values close to unity, where small changes in prevention or deterrent policies would thus move this individuals into the category of ACI.
Expressions \eqref{eq: struct-anom}, \eqref{eq: aci-pred} can be generalised to an alternating renewal process accounting for both time between crimes and time spent in detention.

Altogether, the predictions of ACI have strong implications across all manner of human behaviours. When applied to job durations \cite{employmentdata, jobs}, we encounter attractive companies with high worker retention as employees prefer not to leave. Of the same persistence is encountered in residence patterns \cite{Gordon95, Myers67, Anily99}, this is in indication of popular areas with primarily a sellers' market and rich opportunities for property speculation. Finally, when concerned with criminal activity \cite{crime1, crime2, crime3}, ACI in the recidivism rate indicates individuals or crime types wherein rehabilitation is more likely than re-offending. 
It is expected that similar patterns in behavioural change may be of relevance in marketing for the development of brand loyalty \cite{marketing}, or elite polarisation in political models \cite{elite, elite1}.

One might question the robustness of ACI as described by \eqref{eq: rl-der} and \eqref{eq: struct-anom}. It can be shown that a time-decreasing rate $\gamma(\tau)\sim1/\tau$ arises from population heterogeneity where individuals change behaviour with constant rate, whereupon ACI is a long-term transient phenomenon \cite{jobs}. 
Consequently, each individual need not be subject to (A)CI in order for a company, real estate market or otherwise to reflect that overall trend \cite{bar3}.
Furthermore, anomalous effects can self-organise by the growing popularity of certain positions, neighbourhoods, etc. If the average number of individuals working for a company is $N(t)$, then an increase in $N(t)$ may be indicative of good working conditions there, thus further increasing the number of applicants and popularity of the company. If the rate parameter $\mu(N(t))$ decreases with $N(t)$, then it is possible to have a company which initially is not anomalous, but becomes so as $N$ increases.
The emergence of anomalous effects has previously been studied in \cite{emergence, emergence-dens}.

In this work we have exposed the ubiquitous nature of anomalous cumulative inertia in human activity, supported by empirical evidence from employment, residence, recidivism, and a plethora of other examples. A key contribution is the paradigm shift introduced by recognising the power in anomalous cumulative inertia to predict whether change is likely to occur or not.
As cumulative inertia is an established phenomenon either on the individual or population level, identification of whether a system is anomalous is of fundamental consequence when implementing policy changes. If a system is inertial, but not anomalous, depending on how closely the dynamics resemble ACI (as identified via the parameter $\mu$) we can determine the level of intervention required in order for a certain property to be retained. For example, recidivism times of an ex-convict may follow CI with e.g.\ $\mu=1.1$, whereupon intervention could transfer this individual to ACI-compliant ($\mu<1$) behaviour and thus drastically increase the chances of rehabilitation. This effect can be seen from the graphs of \eqref{eq: eq-dist} and \eqref{eq: struct-anom}. ACI is thus a powerful tool in predicting areas where intervention is likely to be successful by identifying systems close to the boundary $\mu=1$ where changes are likely to have high impact. We underscore this by identifying the resettlement of refugees as one such example.\newline
We believe that the link established in this paper between fractional calculus and the universal phenomenon of ACI provides hitherto unexplored understanding of the human experience as we change homes, jobs and affiliations. This data-driven application of fractional equations for behavioural changes provides an inclusive description of complex systems, which is only reliant on two parameters $\mu,\ \tau_0$.
Future work will apply these tools to particular case studies in order to identify changeable demographic groups, and the suggested targeted areas of policy change.\newline

\textbf{Supplementary Information} is available in the online version of this paper.\newline
\textbf{Author Contributions} HS carried out data analysis. All authors wrote the manuscript.\newline
\textbf{Author Information} Reprints and permissions information is available at \texttt{www.nature.com/reprints}. Correspondence should be addressed to HS (\texttt{helena.stage@manchester.ac.uk}).


\bibliographystyle{unsrt}

\end{multicols}
\end{document}